# Computational limits to the legibility of the imaged human brain

James K. Ruffle FRCR MSc[1], Robert J Gray PhD[1], Samia Mohinta MSc[1], Guilherme Pombo MSc[1], Chaitanya Kaul PhD[2], Harpreet Hyare FRCR PHD[1], Geraint Rees FRCP PhD[1], and Parashkev Nachev FRCP PhD[1]

[1]*Queen Square Institute of Neurology, University College London, London, UK*
[2]*School of Computing Science, University of Glasgow, Glasgow, UK*

Running title:

Computational limits to the legibility of the imaged human brain

Correspondence to:

Dr James K Ruffle

Email: j.ruffle@ucl.ac.uk

Address: Institute of Neurology, UCL, London WC1N 3BG, UK

Correspondence may also be addressed to:

Professor Parashkev Nachev

Email: p.nachev@ucl.ac.uk

Address: Institute of Neurology, UCL, London WC1N 3BG, UK

## Funding

JKR was supported by the Medical Research Council (MR/X00046X/1) and the Guarantors of Brain. PN is supported by the Wellcome Trust (213038/Z/18/Z) and the UCLH NIHR Biomedical Research Centre.

## Conflict of interest

None to declare.

## Manuscript Type

Original article.

## Authorship

Conceptualization: JR, RJG, CK, PN; Data: JR, GP, RJG, GR, PN; Methodology: JR, RJG, SM, GP, CK, PN; Software: JR, RJG, SM, GP, CK; Validation: JR, RJG, SM, GP; Formal analysis: JR, RJG, SM, GP; Manuscript writing JR, PN; Manuscript reviewing, and editing: JR, RJG, SM, GP, CK, HH, GR, PN. All authors have been involved in the writing of the manuscript and have read and approved the final version.

## Data sharing

All code and models are openly available online at https://github.com/high-dimensional/biobank-megamodeller.git. All data is provided courtesy of the UK Biobank – https://www.ukbiobank.ac.uk.

## Acknowledgement

We are grateful to the curators of the UK Biobank for providing this data resource.

## Abbreviations

CNN, convolutional neural network; DWI, diffusion weighted imaging; FDR, false discovery rate; FLAIR, fluid-attenuated inversion recovery; FNN, feed-forward neural network; Hb, haemoglobin; HbA1c, glycated haemoglobin; HDL, high density lipoprotein; LDL, low density lipoprotein; MAE, mean absolute error; MRI, magnetic resonance imaging; rsfMRI, resting-state functional magnetic resonance imaging.

Abstract

Our knowledge of the organisation of the human brain at the *population-level* is yet to translate into power to predict functional differences at the *individual-level,* limiting clinical applications, and casting doubt on the generalisability of inferred mechanisms. It remains unknown whether the difficulty arises from the absence of individuating biological patterns within the brain, or from limited power to access them with the models and compute at our disposal.

Here we comprehensively investigate the resolvability of such patterns with data and compute at unprecedented scale. Across 23 810 unique participants from UK Biobank, we systematically evaluate the predictability of 25 individual biological characteristics, from all available combinations of structural and functional neuroimaging data. Over 4526 GPU*hours of computation, we train, optimize, and evaluate out-of-sample 700 individual predictive models, including fully-connected feed-forward neural networks of demographic, psychological, serological, chronic disease, and functional connectivity characteristics, and both uni- and multi-modal 3D convolutional neural network models of macro- and micro-structural brain imaging.

We find a marked discrepancy between the high predictability of sex (balanced accuracy 99.7%), age (mean absolute error 2.048 years, $R^2$ 0.859), and weight (mean absolute error 2.609Kg, $R^2$ 0.625), for which we set new state-of-the-art performance, and the surprisingly low predictability of other characteristics. Neither structural nor functional imaging predicted an individual's psychology better than the coincidence of common chronic disease ($p<0.05$). Serology predicted chronic disease ($p<0.05$) and was best predicted by it ($p<0.001$), followed by structural neuroimaging ($p<0.05$).

Our findings suggest either more informative imaging or more powerful models will be needed to decipher individual level characteristics from the human brain. We make our models and code openly available.

## Introduction

That the brain exhibits a finely wrought functional-anatomical organisation is no longer in doubt. Macro- and micro-structural features, task-specific and resting state neural activity, focal disruptive and lesion-related neural dependence, all show richly structured, replicable variation across the population[1-11]. But whether these now familiar patterns can explain *individual-level* differences remains an open question[12-15]. Its answer is crucially important for two reasons: first, because the clinical applications of our knowledge of the brain are necessarily addressed not to populations but to individual patients, and second, because the fewer the individuals to which any model generalises, the weaker the grounds it provides for mechanistic inference, no matter how well supported its parameters. It is also a far harder question to address, for individual-level models must inevitably capture the many complex interactions between multiple features on which individual functions may jointly depend. Model architectures of the requisite expressivity[16-18]—whose least upper bound is unknown— plausibly require data of far greater scale and inclusivity than is usual in the field[19], and computational resources rare in neuroscience.

This places us in a Catch 22. Failure to find individually discriminating patterns may be a consequence not of their absence, but of inadequacies of the data and the computational regime[12,20]. Yet calibrating the regime to the demands of the brain's complexity cannot be done with small samples, so distinguishing between the two possibilities is impossible without the resource the distinction is needed to justify in the first place.

How do we break out of this? Models of the necessary complexity here can always be improved upon, so no limit on theoretically achievable fidelity can be definitively set. But we can conduct a *comparative* analysis of a set of biological characteristics, at the current *practical* limit of model expressivity and the compute it requires. If such a state-of-the-art analysis reveals a marked *contrast* of predictability—very high for some characteristics, very low for others— then the conclusion that the unpredictable characteristics are *practically* inaccessible with current data and models is corroborated, and a wholesale change in our approach—data, models, and compute—is motivated (Figure 1). If performance is uniformly poor, then our test may have been inadequate; if it is uniformly excellent, then no change to current practice is indicated.

|                        |      | Variation in predictability |                  |
|------------------------|------|------------------------------|------------------|
|                        |      | *Low*                        | *High*           |
| Maximum predictability | *Low*  | Inadequate test            | Inadequate test  |
|                        | *High* | Maintain status quo        | Regime change    |

Figure 1. A 2x2 factorial relation between the maximum predictability of a set of characteristics and its variation across the set. A failure to achieve high fidelity for any characteristic suggests a general inadequacy of the modelling framework that casts doubt on the quality of the test. Achieving excellent fidelity across all characteristics suggests current approaches are satisfactory. Achieving high fidelity for some features but not for others, suggests a change in the modelling regime is indicated: any or all of data, model, and compute.

Here we conduct such an analysis in a sample of 23 810 unique participants from UK Biobank[1,21], systematically evaluating the predictability of a wide set of individual characteristics from all possible combinations of available neuroimaging data, spanning structural and functional domains. We build and evaluate a suite of 700 discriminative models of different combinations of brain imaging—uni- and multi-modal, macro- and micro-structural, and resting state functional—with biological characteristics ranging across psychology, serology, and disease comorbidity. Over 4526 GPU*hours of computation, including extensive hyper-parameter optimisation, we comprehensively evaluate the individual-level predictability of common biological and pathological characteristics from current brain imaging in this general population cohort. Our analysis sets a new state-of-the-art benchmark for age regression and sex classification (for which we make all model weights open source), demonstrating the felicity of our modelling approach, and reveals a marked heterogeneity of individual predictability that argues for a radical change in the current brain modelling regime.

## Materials and methods

### Data

Data was retrieved from the UK Biobank repository (https://www.ukbiobank.ac.uk)[1,21,22]. From here, we retrieved an unselected fully inclusive representation of the cohort, a sample of 502 505 individuals with 3581 individual variables detailing them. Data missingness of the parameters modelled never exceeded 20% for any variable in our study. We imputed missing variables from this full set using multivariate iterative imputation[23], with hyperparameters of 10000 maximum iterations, a default stopping condition tolerance of 0.001 and an initial strategy of median imputation.

We selected 31 participant variables as predictive modelling targets that engendered a range of domains of individuality. We chose this number as a reasonable balance between high dimensionality of individuating factors, and the volume of models required to be trained for each individual proposed target and anticipating a training time of several months even on cutting-edge computational hardware. We applied an exclusion criterion after variable selection, where if a feature was categorical/binary in nature, it was excluded if an imbalance between majority and minority class were 10:1 or greater. The rationale was to minimize the impact of imbalance on our evaluations of predictability. This led to 3 variables being removed (namely pathological classes of previous stroke, myocardial infarction, and the presence of type 2 diabetes), leaving 28 variables for possible targets.

Next, we computed the pairwise Maximal Information Coefficient[24] between all 28 targets. The purpose of this was to identify features that were highly collinear to one another, limiting the interpretability of their individual prediction where they are jointly modelled. This led to the removal of 3 further variables, namely neuroticism (closely related to several other psychological factors, haematocrit (closely related to haemoglobin concentration), and body mass index (closely related to weight). In total, this process yielded 25 unique and minimally inter-related target features (Figure 2, Supplementary Figure 1, Supplementary Table 1). The choice to not select further variables was entirely driven by available computing resource and the substantial GPU-hours required for the large number of possible models with different input features to predict each target.

We then organised these 25 targets into their respective domains, as follows: i) Constitutional, comprising sex, age, weight and handedness; ii) Psychology, comprising feelings of guilt, loneliness, worry, feel tense, anxious, nervous, fed-up, sensitive, irritable, miserable, mood-swings, and reaction time (ms); iii) Disease, comprising body fat (obesity), hypertension, asthma, atopy, smoking (addiction); and iv) Serological, comprising concentrations of haemoglobin (g/dl), HbA1c (mmol/mol), HDL (mmol/L) and LDL (mmol/L) (Figure 2, Supplementary Figure 1, Supplementary Table 1).

Participant selection

Our next task was to delineate the participants who had undergone comprehensive neuroimaging investigations inclusive of T1-weighted (T1), fluid-attenuated and inversion recovery (FLAIR), diffusion-weighted imaging (DWI), and functional magnetic resonance imaging (fMRI) sequences. Brain MRI data was available in our server for the following: FLAIR (n=39 276), T1 (n=34 041), DWI (n=38 909), and fMRI (n=31 748). For participants with multiple imaging attendances, we utilised only the first MRI study to prevent an information leak. We removed 8 participants with artefact degraded MRI sequences and subsampled the cohort to include those with all four imaging dimensions of MRI data (Figure 2). This yielded a suitable sample of 23 810 unique participants (11 141 male, 12 669 female, mean age ± standard deviation (SD) 54.775 ± 7.44 years). A breakdown of this data is provided as Table 1.

Pre-processed brain images were used from the pipeline as curated by the brain imaging leads to the UK Biobank, as described elsewhere[21]. All data were held in compressed NIFTI format. Formulations of these imaging sequences included the original raw data and imaging registered to Montreal Neurological Institute (MNI) template space. This pipeline relied on the validated toolkit developed by the FMRIB Oxford team and the FMRIB software library (FSL)[25]. This included brain extraction[26], and image registration[27] to the MNI152 nonlinear sixth generation standard space T1 brain template[28]. We utilised the Biobank-release pre-processed 3-dimensional volumetric T1-weighted and FLAIR structural acquisitions, the tract based spatial statistics (TBSS) pre-processed fractional anisotropy (FA) maps, and the pre-processed 4-dimensional (4D) volumetric blood oxygen level dependent (BOLD) fMRI acquisition[21,29]. From the pre-processed 4D time-series fMRI, we derived z-scored standardized functional connectivity matrices[30] in accordance with the Glasser cortical parcellation scheme, comprising 360 regions of interest (180 per hemisphere)[31]. With pairwise connectivity this yields 64 620 unique edges. We did not use thresholding on the edges, for the downstream deep learning architectures would undertake feature selection intrinsically.

The rationale to reduce dimensionality of 4D time-series fMRI data to a connectome representation was twofold: firstly, since connectivity analyses now form a mainstay of neuroimaging analysis in this scientific domain[32-34]; and secondly, to reduce computational demand that would otherwise be infeasibly large with direct modelling of the raw fMRI time-series in what would otherwise demand a 4D convolutional neural network, infeasible for this study in scale of sample size, algorithm (GPU size), and the number of models to train. We deliberately chose not to use the mean BOLD 3D image across the 4D timeseries (using instead the aforementioned connectivity approach), as the single mean 3D image would not capture variances in haemodynamic response, and therefore was felt likely to be reductive[35]. Moreover, it seemed prudent to use data formulations like that which is used across the bulk of neuroimaging research, which could form helpful benchmarks in model fidelity. Similarly, we deliberately chose not to pass other measures of structural connectivity, such as gray matter similarity[36] or tractography[34,37,38], as it was felt likely these data would at least in part be leveraged from the 3D T1/FLAIR and DWI images, respectively.

Sample partitioning

We partitioned the suitable cohort of 23 810 unique participants into training, evaluation and testing sets, using 80% of samples for training (n =19 048), 10% for model validation (n = 2381) and reserving the remaining 10% (n = 2381) for model testing on completely unseen data (Figure 2). This data partition was performed prior to any modelling, with the precise partition maintained for *every* experiment undertaken to ensure their comparability.

We statistically compared all modelling targets across the training, validation, and testing partitions. For continuous targets, these were compared with one-way analysis of variance (ANOVA), and for categorical targets, Chi-squared. P values were corrected for multiple comparisons by False Discovery Rate using the Benjamini-Hochberg method (Table 1)[39].

## Bayesian graph representations of interactions amongst non-imaging features

To investigate the pairwise relations between the 25 variables we employed graph analysis of the training sample non-imaging data. Given the variety of variable types—continuous or categorical—we used mutual information as the primary index of similarity. This was constructed into the format of an undirected graph, wherein nodes were the targets of study, and the edges were mutual information between pairwise features – creating a graph of 25 nodes and 300 edges. The weighted eigenvector centrality of nodes was also calculated. From these data, we fitted a layered nested stochastic block model, a generative model of the community structure of graphs[40,41], passing the association direction as a layer property. We further equilibrate the stochastic block model fit with Markov-Chain Monte Carlo (MCMC) simulated annealing to optimise the partition in accordance with minimizing the description length entropy criterion, as detailed elsewhere[42-44] (Figure 2).

## Approach

### Multi-modal modelling of volume brain imaging data

We modelled T1 and FLAIR sequences for macrostructure, DWI FA for microstructure, and functional connectivity matrices derived from participant BOLD timeseries for resting state function. The non-target (for each experiment) non-imaging data were organised across the domains of constitutional, psychology, serology, and disease. It is over these seven (three imaging, four non-imaging data) domains that we could evaluate multimodal performance.

### Combinatorial analysis of the imaging and target feature space

Having identified a set of unique targets and predictors from a large population, we set out to perform a systematic combinatorial analysis to determine what targets could—and just as importantly could not—be predicted by machine models, from predictors taken alone or in combination.

First, we constructed models to predict targets within the constitutional domain, i.e., participant sex, age, handedness, and weight. We examined all combinations of the imaging modalities for this case out of i) T1 + FLAIR, ii) DWI, iii) rsfMRI connectivity, and all combinations of the former, i.e., the largest model feature set would therefore be T1 + FLAIR + DWI + rsfMRI connectivity. This yielded 7 different models to fit for each of the 4 constitutional targets. These models would also serve as a benchmark to quantify the felicity of architectural choices in comparison with the extant literature.

We applied a similar approach to the targets across the psychology, serology, and disease domains, fitting models with the same imaging combinations. Constitutional data is typically collected as a standard part of a research experiment (including in neuroimaging), and used for either predictive modelling, nuisance covariates, or even as the variable of interest. To that end, we decided the available inclusion of these constitutional features to all other models of these non-constitutional targets was also reasonable. Similarly, we quantified the benefit of providing further non-imaging data from domains different to the current target, which further supplemented the number of possible imaging and non-imaging input feature sets for a given target significantly (i.e., 32 unique combinations for each target).

Overall, this process yielded 28 unique models to be trained across the constitutional targets (7 combinations * 4 targets = 28 models), and 672 unique models across the psychology, serology, and disease targets (32 combinations * 21 targets = 672 models). This yielded a total of 700 models to be trained in this experimental design.

Algorithm

*A role for complex models*

A typical volume brain image is a 182 x 218 x 182 matrix of voxels: *more than 7 million variables*. This is for a *single* – unimodal - imaging sequence. Our task is to capture complex biological and pathological traits about individuals from high-dimensional data, a task necessarily best solved by models of sufficient complexity to capture this richness. 3D convolutional neural networks (CNNs) offer a potential solution to this problem and have become state-of-the-art for modelling brain imaging data across numerous tasks[45-48].

*Remediating target class imbalances*

Class imbalance was handled by randomly sub-sampling the majority class, performed at the beginning of each epoch. No under- or over-sampling was applied to the validation of test partitions, but performance metrics were always balanced to accommodate class imbalances.

*Data pre-processing and augmentation*

For each model target, continuous variables were clamped between the $0.5^{th}$ and $99.5^{th}$ percentile, z-scored, and normalized to the range -1 to 1. The targets this applied to were age, weight, haemoglobin, reaction time, body fat, HBA1c, HDL, and LDL. Categorical targets were one-hot encoded. The reasoning behind re-scaling continuous targets into a -1 to 1 range space (for example, as opposed to modelling age in years), was so that all models across different continuous targets with different native ranges were more optimally comparable in both loss function and evaluation metric.

For associated non-imaging data (e.g., any combination of constitutional, psychology, serology, and disease features), the selected combination was first passed to the code, with any unused non-imaging data (including that of the same block to the model target) zeroed. For example, in training a model to predict anxiety, using constitutional and serological non-imaging data, all other psychological data would be zeroed since anxiety were within this target domain, but disease data would also be zeroed as it was not selected to be passed to the model. The reasoning behind this was to 1) ensure the prevention of an information leak between similar feature domains (e.g., it seemed probable one could fit a function of anxiety from a selection of other measures of psychology), but 2) would maintain precisely the same modelling architectural complexity, only where some features were encoded as zero.

For structural neuroimaging, we developed a comprehensive MRI augmentation pipeline using MONAI[49,50]. This pipeline included the following: image resizing; ii) clamping along the 0.5th and 99.5th percentile; iii) intensity normalisation; intensity scaling to the range of -1 to 1; iv) random histogram shifting; v) random intensity scaling; vi) random affine transformations; vii) random 3D elastic deformations; vii) image re-normalization and viii) image re-scaling to the range of -1 to 1. All random transformations were with a probability of application of 0.1.

To evaluate the discrepancy (if any) in model fidelity at differential imaging resolutions, we considered two image resolutions to resize data to. We firstly fitted the 700 models with resizing images from native 182 x 218 x 182, isotropic and volumetric, 1mm$^3$ voxel dimensions to a smaller 64 x 64 x 64 isotropic resolution. Doing so would enable models to train significantly faster and illuminate where performant signal for a given target could be extracted from the structural 3D sequences. This enabled training one model on four Tesla P100 16Gb GPUs within a DGX environment with a batch size of 64. Having trained these models, we identified the best sets of input combinations for each of the 25 targets and retrained each (i.e., the best model, per target), with resizing images from native 182 x 218 x 182, isotropic and volumetric, 1mm$^3$ voxel dimensions to a 128 x 128 x 128 isotropic resolution. This enabled training one model on eight Tesla P100 16Gb GPUs within a DGX environment with a batch size of 32.

Like the associated non-imaging data pipeline, where a given MRI sequence was not chosen as a data input to a given target, it was zeroed. For example, when training a model to predict sex with T1+FLAIR, the DWI channel was zeroed. The rationale for this was to maintain the same model size/architecture regardless of the data passed to it.

For functional connectivity, we used the bilateral Glasser parcellation of 360 regions (180 per hemisphere) to extract a given regions BOLD time-series signal, and cross-correlate to a symmetrical adjacency matrix of shape 360 x 360, using nilearn[30]. Data were standardised by z-score transformation. For each participant, the upper triangle of this symmetrical rsfMRI connectivity matrix was extracted and flattened to a 1D array, with 64 620 functional connections between each pairwise set of regions. Like the remaining pipeline, where connectivity was not selected as an input to a given model, it was zeroed. We opted for the Glasser parcellation scheme[31] since it is one of the most widely used and cited (3965 tracked Google Scholar citations as of 01/03/2024). We did consider the use of other functional templates to boost the analysis further, however since the computing requirements for all modelling was already substantial, we did not pursue it further. This could however be explored in future research.

Architecture

All deep learning aspects of the study were undertaken using PyTorch[51] and MONAI[49], with model architectures as listed below.

*Feed-forward neural network for non-imaging data*

For modelling with non-imaging data, we constructed a feed-forward neural network (FNN), which took an input dimension of 24, the number of non-imaging data features minus 1 (the target), with sequential dense layers of 128, 64 and 32 with sequential batch normalisation[52], Gaussian error linear unit (GELU) activation[53], and dropout (rate 0.1)[54]. The non-imaging data FNN comprised 13 984 parameters.

*Three-dimensional convolutional neural networks for volumetric imaging data*

We developed a 3D convolutional neural network (CNN) architecture for modelling with volume brain imaging data[17,55]. This CNN was contained three channels, for T1, FLAIR and DWI. The architecture followed the sequence of 3D convolution, GELU hidden activation[53], skip convolution layers[56], batch normalization[52], GELU hidden activation[53], max 3D pooling[57], dropout (rate 0.1)[54], flattening to a linear dense layer, batch normalisation[52], and GELU output activation[53]. For training with 64 x 64 x 64 images, the CNN channel sizes were 32, 64, 128, 256, 256 with a further final output channel of size 128. For training with 128 x 128 x 128 images, the CNN channel sizes were 32, 64, 128, 256, 256, 256 with a further final output channel of size 256. Our architectural design and channel sizes were guided by review of existing literature, and benchmark comparisons to open-source datasets inclusive of MNIST and MNIST-fashion[56,58,59]. Skip convolutions have been shown advantageous to well-known models such as that of ResNet, with a building block composed of two convolution layers and activation operators, then concatenated[56,60]. The CNN comprised 7 427 680 parameters for 64 x 64 x 64 resolution models, and 11 295 968 for 128 x 128 x 128 models.

*Feed-forward neural network for functional connectivity*

For modelling with functional connectivity, we constructed a FNN which took an input dimension of 64 620, the number of unique pairwise functional connections in the flattened connectivity matrix, with sequential dense layers of 128 and 128, each with batch normalisation[52], GELU activation[53], and dropout (rate 0.5)[54]. This rsfMRI connectivity FNN comprised 8 288 512 parameters.

*Model concatenation and feed-forward neural network for final prediction*

Outputs from the non-imaging data FNN, multi-channel 3D-CNN connectivity, and rsfMRI connectivity FNN, were concatenated and used for a final FNN for target prediction. This took the sum of the output channels from above (288 when training with 64 x 64 x 64 MRI, 416 when training with 128 x 128 x 128), with further dense layers of size 256, 256, and a final layer of size 2 where the target was categorical and one-hot encoded (e.g., sex), or 1 where the target was continuous (e.g., age). Similar to the non-imaging data and connectivity FNN models, these used sequential batch normalisation[52], GELU activation[53], and dropout (rate 0.1)[54]. When training with an MRI resolution of 64 x 64 x 64, this final FNN comprised 141 057 parameters, and 173 825 when training in 128 x 128 x 128.

The total parameter count for this multi-dimensional modelling architecture was 15 871 233 when training with an MRI resolution of 64 x 64 x 64, and 19 772 289 when training with an MRI resolution of 128 x 182 x 128. The full modelling architecture can be visualised in Supplementary Figure 2.

*Hyperparameters*

All models were compiled with a learning rate of 0.0001, the Adam optimiser[61], $L_2$ regularisation[62], a batch size of 64 for 64 x 64 x 64 resolution MRI models, and 32 for 128 x 128 x 128 (limited only by GPU size). Models were permitted to train for anywhere up to 100 000 epochs, but with early stopping if there was failure to improve the validation loss function after 50 epochs. For categorical targets, the loss function was binary cross entropy, and for continuous, mean squared error.

Model evaluation

Models were always trained on the training data only and evaluated at the end of each epoch with the validation set. The best performing model, criterion on the loss function, were saved. After completion of model training, the best performing epoch for each model (on validation data) were used for evaluation of performance on the completely unseen test data. Numerous performance metrics were derived, including accuracy, precision, recall, F1, a confusion matrix[63], the receiver operator characteristic (ROC) curve, the $R^2$ and r for continuous targets, the model loss, and the amount of time each model required to train[64,65]. Metrics of categorical performance were always balanced by macro average to accommodate for any degree of class imbalance. It should be noted that we place our focus here on evaluating categorical models with metrics appropriate for any degree of class imbalance, such as with balanced accuracy or macro-averaged precision, recall, and F1. However, we also provide AUROC as there are a high proportion of research articles that only report this despite its clear limitation to sample imbalance, such that our work is still comparable to the broader literature. To enable large-scale comparison across all 700 models, irrespective of if a categorical or continuous target, we also converted the task to categorical with continuous targets divided by median split with respect to the training dataset. We determined the 'best' performing model by the highest balanced accuracy in the testing set for categorical target, and by the highest $R^2$ for continuous targets[64,65].

Model validation with open-source datasets

To ensure that the performance of our CNN model architectures was comparable to existing state-of-the-art performances, we prototyped CNNs initial with both MNIST and the MNIST-fashion (data not shown).

## Model comparison

*Linear mixed-effects models*

After fitting all possible model combinations, we undertook post-hoc comparisons of models, reviewing the performance metrics to identify both human factors inherently predictable by imaging data, but also the value of data components in fitting each factor. This was undertaken by visual inspection of all performance metrics, including ROC curves and confusion matrices.

We conducted formal statistical comparison of model performance with linear mixed-effect models[66,67]. These were in the following formulation:

$$Model\ performance \sim T1 + FLAIR + DWI + rsfMRI\ Connectivity +$$
$$Psychology + Disease + Serology + (1\ |\ Target),$$

where for constitutional, psychological and disease targets, model performance were by the balanced accuracy given the majority were categorical in formulation, but included conversion of continuous targets by median split, as described above. Whereas for serology targets, model performance was $R^2$.

*Graph representations*

We derived graph representations of model performances across the 25 targets, across all possible feature combinations. This was undertaken by fitting an undirected graph of all targets as nodes, wherein the edge weights were calculated as the inverse of the Euclidean distance between all performance metrics for a given model[68]. We also used this data to derive the weighted eigenvector centrality of each node, weighted by the similarity across different performance metrics, that might further provide insight in explaining the similarities and dissimilarities across model performances. Lastly, we converted these results into a fully interactive HTML object[69] to enable reader visualisation.

## Compute

*Hardware*

Local development and prototyping were predominantly performed on a 32 core (64 thread) CPU Linux workstation housing 135Gb of RAM and an NVIDIA 2080Ti GPU (11Gb size), OS Ubuntu 20.04. All model training was undertaken on a DGX workstation housing 8 x P100 16Gb GPUs, 80 CPU threads and 503Gb of RAM.

*Software*

Most of the programming was undertaken within a Python environment (version 3.6.9). Further small operations were completed with Bash for faster IO enabled by GNU parallel[70].

The following Python packages were utilised: graph-tool[68], gravis[69], matplotlib[71], minepy[72], MONAI[49], nibabel[73], NumPy[74], pandas[75], PyTorch[51], scikit-learn[23], SciPy[76], seaborn[77], statsmodels[78]. GPU-modelling was achieved with the CUDA toolkit version 11.0[79]. Linear mixed effect models were constructed within R version 4.1.2, using packages Tidyverse[67], and nlme[66].

### Ethical approval

The study was approved by local institutional review board and conducted in accordance with the "Declaration of Helsinki". Use of UK Biobank data were approved under study application identifier #16273.

### Code, model, and data availability

All code and models are openly available online at https://github.com/high-dimensional/biobank-megamodeller.git. Supplementary code evaluating age and sex prediction using other models[80-82] is also available. All data is available from the UK Biobank curators.

## Results

### Cohort

We studied an unselected sample of 23 810 unique UK Biobank participants (11 141 male, 12 669 female, mean age ± 95% confidence interval (CI) 54.775 years (54.66 – 54.85)) who underwent multi-sequence MRI, including T1-weighted, fluid-attenuated inversion recovery (FLAIR), diffusion weighted imaging (DWI), and resting-state functional MRI (rsfMRI) acquisitions, and for whom a set of 25 constitutional, psychological, disease, and serological domain variables of plausible clinical or scientific interest was available. A compact set of characteristics was chosen to enable comprehensive modelling of the comparative predictability of subsets of variables from the remainder, a task that scales exponentially with the number of subsets. Participants were randomly partitioned into training, validation, and testing sets of 19 048, 2381, and 2381 unique participants, respectively, with no significant differences between them (Figure 2, Table 1).

| Target (Domain) | Training (n=19 048) | Validation (n=2381) | Testing (n=2381) | FDR-P value | Statistical test |
|---|---|---|---|---|---|
| Sex (C) | Female (n=10163), Male (n=8885) | Female (n=1267), Male (n=1114) | Female (n=1239), Male (n=1142) | 0.64 | Chi-square |
| Age (C) | 54.74 (54.64 - 54.85) | 54.59 (54.29 - 54.89) | 55.01 (54.71 - 55.3) | 0.45 | ANOVA |
| Weight (C) | 76.5 (76.3 - 76.71) | 76.45 (75.86 - 77.03) | 77.22 (76.62 - 77.81) | 0.35 | ANOVA |
| Handedness (C) | Right (n=16983), Left (n=2065) | Right (n=2101), Left (n=280) | Right (n=2116), Left (n=265) | 0.57 | Chi-square |
| Mood Swings (P) | No (n=11252), Yes (n=7796) | No (n=1415), Yes (n=966) | No (n=1370), Yes (n=1011) | 0.55 | Chi-square |
| Miserableness (P) | No (n=11306), Yes (n=7742) | No (n=1415), Yes (n=966) | No (n=1354), Yes (n=1027) | 0.35 | Chi-square |
| Irritability (P) | No (n=13804), Yes (n=5244) | No (n=1760), Yes (n=621) | No (n=1681), Yes (n=700) | 0.34 | Chi-square |
| Sensitivity (P) | No (n=8998), Yes (n=10050) | No (n=1155), Yes (n=1226) | No (n=1094), Yes (n=1287) | 0.48 | Chi-square |
| Fed Up (P) | No (n=12287), Yes (n=6761) | No (n=1504), Yes (n=877) | No (n=1479), Yes (n=902) | 0.34 | Chi-square |
| Nervous (P) | No (n=15288), Yes (n=3760) | No (n=1917), Yes (n=464) | No (n=1913), Yes (n=468) | 0.99 | Chi-square |
| Anxious (P) | No (n=8953), Yes (n=10095) | No (n=1116), Yes (n=1265) | No (n=1120), Yes (n=1261) | 0.99 | Chi-square |
| Tense (P) | No (n=16320), Yes (n=2728) | No (n=2013), Yes (n=368) | No (n=2022), Yes (n=359) | 0.48 | Chi-square |
| Worry (P) | No (n=9860), Yes (n=9188) | No (n=1225), Yes (n=1156) | No (n=1207), Yes (n=1174) | 0.76 | Chi-square |
| Lonely (P) | No (n=16245), Yes (n=2803) | No (n=2033), Yes (n=348) | No (n=2027), Yes (n=354) | 0.99 | Chi-square |
| Guilty (P) | No (n=13820), Yes (n=5228) | No (n=1701), Yes (n=680) | No (n=1714), Yes (n=667) | 0.64 | Chi-square |
| Reaction Time (ms) (P) | 537.17 (535.78 - 538.57) | 534.12 (530.34 - 537.91) | 536.45 (532.54 - 540.35) | 0.56 | ANOVA |
| Smoking (D) | No (n=11670), Yes (n=7378) | No (n=1482), Yes (n=899) | No (n=1456), Yes (n=925) | 0.76 | Chi-square |
| Hypertension (D) | No (n=15612), Yes (n=3436) | No (n=1943), Yes (n=438) | No (n=1940), Yes (n=441) | 0.89 | Chi-square |
| Atopy (D) | No (n=14736), Yes (n=4312) | No (n=1807), Yes (n=574) | No (n=1812), Yes (n=569) | 0.45 | Chi-square |
| Asthma (D) | No (n=17210), Yes (n=1838) | No (n=2135), Yes (n=246) | No (n=2131), Yes (n=250) | 0.52 | Chi-square |
| Body Fat (%) (D) | 30.1 (29.99 - 30.22) | 30.03 (29.71 - 30.36) | 30.24 (29.91 - 30.57) | 0.76 | ANOVA |
| Hb (g/dl) (S) | 14.16 (14.15 - 14.18) | 14.17 (14.12 - 14.22) | 14.21 (14.16 - 14.26) | 0.48 | ANOVA |
| HbA1c (mmol/mol) (S) | 34.92 (34.86 - 34.99) | 35.03 (34.85 - 35.22) | 35.06 (34.88 - 35.23) | 0.48 | ANOVA |
| HDL (mmol/L) (S) | 1.48 (1.47 - 1.48) | 1.48 (1.46 - 1.49) | 1.46 (1.45 - 1.47) | 0.34 | ANOVA |
| LDL (mmol/L) (S) | 3.59 (3.58 - 3.6) | 3.57 (3.54 - 3.61) | 3.61 (3.57 - 3.64) | 0.56 | ANOVA |

**Table 1**: Cohort statistics across all modelled targets within training, validation, and testing data partitions. The domain of each target is demarcated by the bracketed letter: C, constitutional; P, psychology; D, disease; S, serology. Categorical features are shown with the number of entries to each category, whilst continuous are shown with mean and 95% confidence interval. Statistical testing across training, validation, and testing sets shows there are no significant cohort differences between each partition. All p values are False-Discovery-Rate corrected.

## Graph community representations of non-imaging characteristics

A generative stochastic block model[83], with separate layers[40] for positively and negatively related characteristics across the four non-imaging domains, was used to derive a succinct hierarchical representation of the non-imaging data in terms of its patterns of distinct covariance, captured by the graph model as hypergraph 'community' structure. The largest community consisted of psychological measures. The second community consisted of sex, age, haemoglobin (Hb), weight, percentage body fat, and high-density lipoprotein (HDL). Male gender was associated with haemoglobin (mutual information (MI) 0.30) and weight (MI 0.18), and inversely related to with body fat (MI 0.31) and HDL (0.14). The final community, consisting of low-density lipoprotein (LDL), HbA1c, reaction time, asthma, atopy, hypertension, smoking, and handedness, was characterised by weak mutual information across all features (Figure 2). A stochastic block model employing a non-linear index of dependence based on the maximal information coefficient[24], revealed essentially the same structure (Supplementary Figure 1).

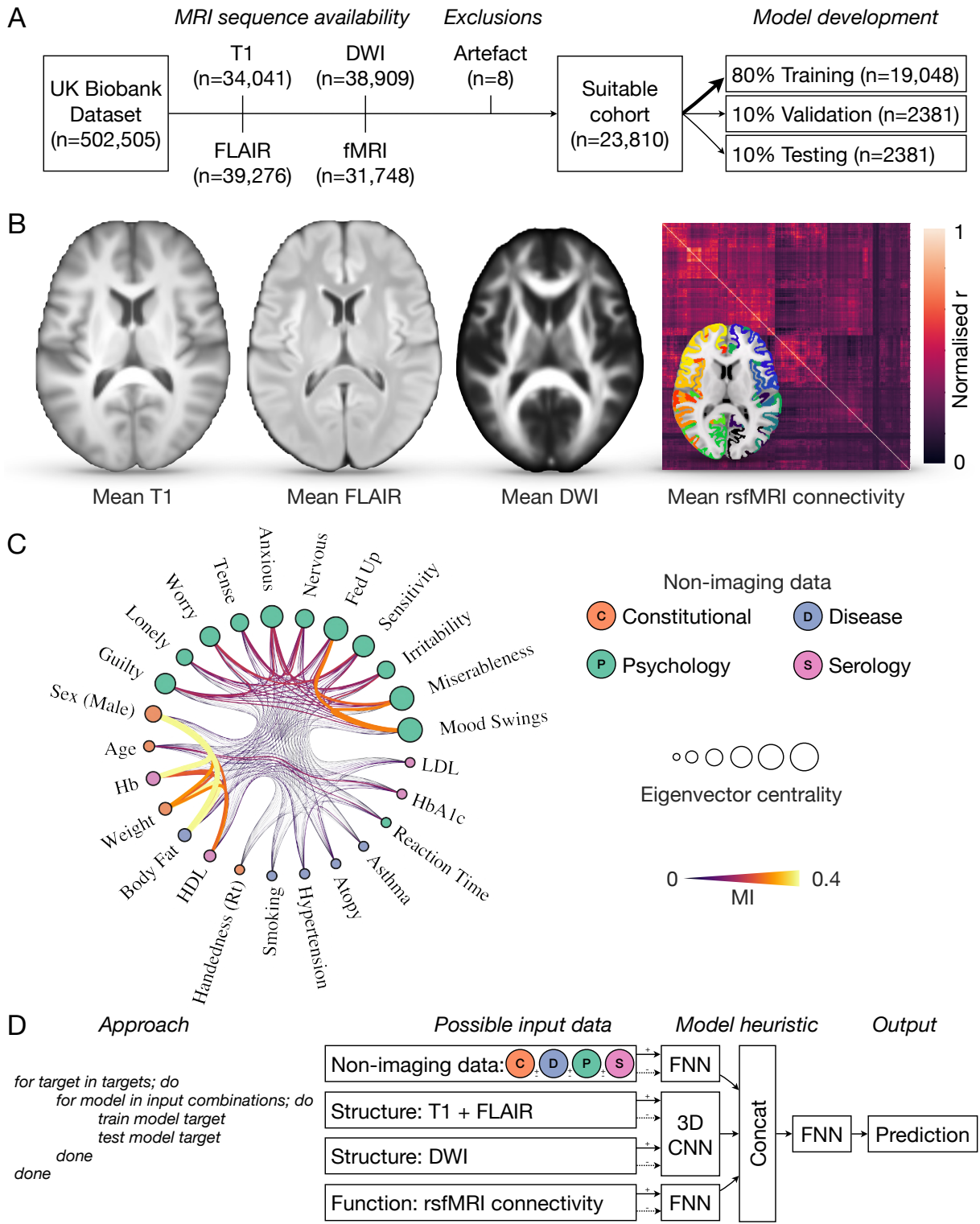

Figure 2: Approach. A) Data selection and partitioning. B) Mean T1-weighted, FLAIR, DWI images, and rsfMRI connectivity matrix across the full cohort of 23 810 participants. C) Layered, nested, generative stochastic block model of modelling targets, with edges depicting the strength of interconnection by mutual information. Node size is proportional to eigenvector centrality, a measure of node 'influence' across its network. D) Algorithmic approach for exploring the model target-feature space to distinguish targets that can be reliably predicted from those that cannot, across all possible data inputs. Shown here is also a schematic of the possible data to train with, ranging from non-imaging data across the constitutional (C) – orange, disease (D) – blue, psychology (P) – green, and serology (S) – pink feature domains; and T1/FLAIR volumetric structural imaging; DWI volumetric imaging; and rsfMRI connectivity. These data are passed to individual trainable model blocks: a fully-connected feed-forward

network (FNN) for both non-imaging data, and rsfMRI connectivity, and a 3D convolutional neural network (CNN) for T1/FLAIR and/or DWI. Model block dense layers are then concatenated and passed to a final FNN for output prediction.

## Imaging models of constitutional characteristics achieve state-of-the-art performance

Across constitutional characteristics, models of FLAIR, T1, and DWI achieved state-of-the-art sex classification (balanced accuracy (BA) 99.7%, area under the receiver operator characteristic curve (AUROC) > 0.999); age regression ($R^2$ of 0.859, mean absolute error (MAE) of 2.048 years); and weight regression ($R^2$ of 0.625, MAE of 7.042kg) performance (Figure 3). Models of rsfMRI connectivity alone performed the worst on these characteristics but yielded the best prediction of handedness (BA 57.7%, AUROC of 0.915). Our brain age model outperforms previous top prediction models including that of Peng et al.,[81] (MAE reported in manuscript=2.14 years, MAE from evaluation on our test set=5.282 years), and Cole et al.,[80] (MAE reported in manuscript=3.55 years, MAE from evaluation on our test set=5.115 years). Similarly, our sex classifier outperformed previous top prediction model from Peng et al.,[81] (accuracy reported in manuscript=99.5%, accuracy from evaluation on our test set=96.6%).

To quantify the relative contribution of each imaging feature we employed a linear mixed-effect model predicting balanced accuracy from the choice of imaging inputs. This showed the inclusion of T1/FLAIR structural sequences to be significantly advantageous to model performance (coefficient 0.041, 95% CI 0.012 to 0.0694, p=0.008) (Figure 4). There were non-significant trends for the inclusion of both DWI (coefficient 0.0211, 95% CI -0.007 to 0.050, p=0.138) and rsfMRI connectivity (coefficient 0.012, 95% CI -0.017 to 0.040, p=0.408).

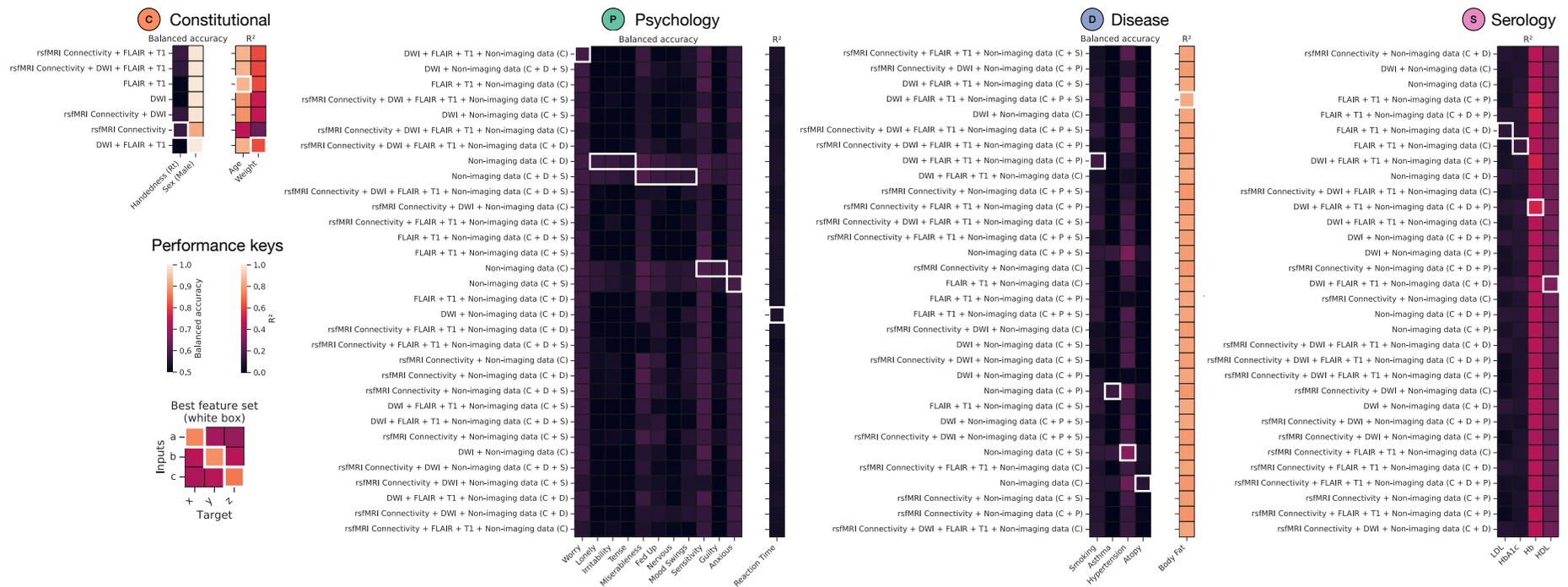

Figure 3: Model performances. Test set performance for all models across constitutional (C) – orange, disease (D) – blue, psychology (P) – green, and serology (S) – pink feature domains. Index of performance is given as balanced accuracy for classification targets and $R^2$ for regression fits. The x-axis of all heatmaps depicts the model target, and y-axis depicts the range of feature inputs. White boxes demarcate the best set of inputs to achieve the greatest out-of-sample model performance.

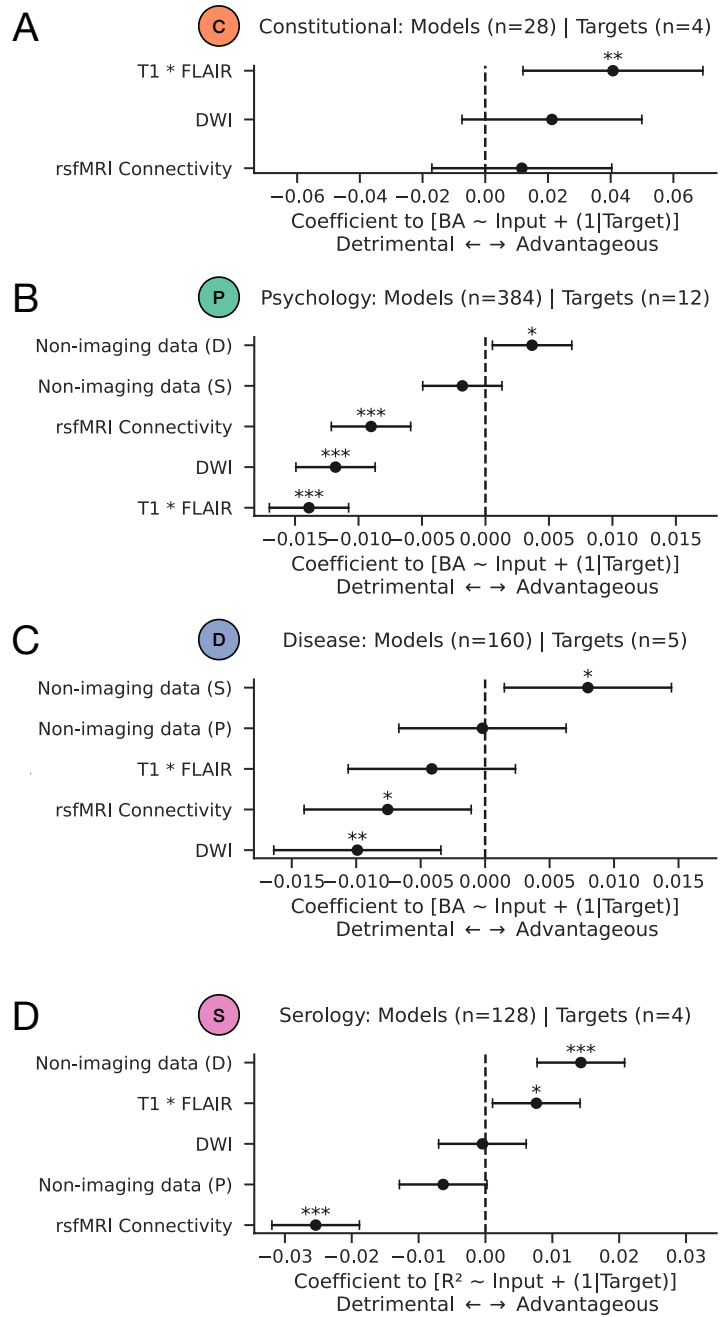

**Figure 4: Domain-specific effects.** Linear mixed-effects models for predicting out of sample performance (balanced accuracy or $R^2$, where applicable) from structural imaging, functional imaging, and non-imaging domain feature sets. Shown are coefficient plots for models whose targets are A) constitutional, B) psychology, C) disease, and D) serology. Inputs with coefficients whose values are positive are associated with increase model performance (advantageous), whilst features with negative coefficients are associated with weaker performance (detrimental). Asterisks stipulate statistical significance as per standard convention: * denotes p<0.05; ** denotes p<0.01; *** denotes p<0.001.

## Psychological characteristics are poorly predicted by imaging

Models predicting psychological characteristics exhibited poor test-set performance (Figure 3), in the face of extensive optimisation. Models of non-imaging data *alone,* offered the best prediction fidelity in 10 of the 12 psychological characteristics. Specifically, constitutional non-imaging data *alone* best predicted sensitivity (BA 60.1%, AUROC 0.614), and guilt (BA 58.1%, AUROC 0.659). Models of constitutional and disease non-imaging data best predicted loneliness (BA 57.5%, AUROC 0.685), irritability (BA 57.5%, AUROC 0.628), and the propensity to feel tense (BA 55.8%, AUROC 0.667). Constitutional and serological non-imaging data best predicted anxiety (BA 59.0%, AUROC 0.607). Lastly, models featuring constitutional, serological, and disease non-imaging data best predicted nervousness (BA 57.3%, AUROC 0.650), feeling fed up (BA 57.7%, AUROC 0.649), miserableness (BA 60.5%, AUROC 0.650), and mood swings (balanced accuracy 56.6%, AUROC 0.611). Indeed, only the propensity to worry and reaction time were best predicted with the inclusion of any neuroimaging data. Worry was best predicted with DWI, FLAIR, T1, and constitutional non-imaging data (BA 58.8%, AUROC 0.601), whereas reaction time was best predicted with DWI, constitutional and disease non-imaging data ($R^2$ of 0.081, MAE of 70.316ms).

A linear mixed-effect model predicting balanced accuracy from the choice of T1/FLAIR, DWI, rsfMRI connectivity, serology, and disease data for all 32 model input combinations for the 12 psychology targets (384 models) found disease significantly advantageous to model performance (coefficient 0.004, 95% CI 0.001 to 0.007, p=0.021) (Figure 4). The inclusion of T1/FLAIR (coefficient -0.014, 95% CI -0.017 to -0.011, p<0.001), DWI (coefficient -0.012, 95% CI -0.015 to -0.009, p<0.001) and rsfMRI connectivity (coefficient -0.009, 95% CI -0.012 to -0.006, p<0.001) were all significantly associated with detrimental model performance. There was no significant relationship between serological non-imaging data and model performance (p=0.252).

## Disease characteristics are best predicted by serological data, not neuroimaging

Test performance for models predicting individual disease characteristics was more variable, revealing 3 of 5 disease targets to be best predicted from non-imaging data alone. These were: 1) a diagnosis of atopy, using constitutional non-imaging data alone (BA 54.4%, AUROC 0.656), 2) asthma, using constitutional and psychological non-imaging data (BA 57.8%, AUROC 0.608), and 3) hypertension, using constitutional and serological non-imaging data (BA 66.7%, AUROC 0.774). Smoking was best predicted by DWI, FLAIR, T1, constitutional and psychological non-imaging data (balanced accuracy 58.1%, AUROC 0.682), and percentage body fat using DWI, FLAIR, T1, constitutional, psychological, and serological non-imaging data ($R^2$ of 0.834, r of 0.914, and MAE of 2.653%) (Figure 3).

A linear mixed-effect model predicting balanced accuracy from the inclusion of T1/FLAIR, DWI, rsfMRI connectivity, serology, and psychology data for all 32 model input combinations for the 5 disease targets (160 models) found serology significantly advantageous to model performance (coefficient 0.008, 95% CI 0.002 to 0.015, p=0.016) (Figure 4). The inclusion of DWI (coefficient -0.010, 95% CI -0.016 to -0.003, p=0.003) and rsfMRI connectivity (coefficient -0.008, 95% CI -0.014 to -0.001, p=0.023) were significantly associated with detrimental model performance. There was a non-significant trend for the inclusion of T1/FLAIR imaging to also be detrimental (coefficient -0.004, 95% CI -0.011 to 0.002, p=0.209). There was no significant relationship between psychology non-imaging data and model performance (p=0.950).

Models of serology perform best with multi-modal imaging and non-imaging data

Serological targets were best predicted from variable combinations of imaging and non-imaging data. The best performing haemoglobin model included FLAIR, T1, and DWI sequences, with constitutional, psychological, and disease non-imaging data, achieving an $R^2$ of 0.524, r of 0.725, and MAE of 0.629g/dl. HDL was best predicted by FLAIR, T1, and DWI sequences, augmented with both constitutional and disease non-imaging data, achieving an $R^2$ of 0.309, r of 0.556, and MAE of 0.209mmol/L. Prediction of HbA1c was weaker, the best feature combination of which was FLAIR, T1, and constitutional non-imaging data, achieving an $R^2$ of 0.146, r of 0.394, and MAE of 2.790mmol/mol. LDL concentration was similarly weak, though the best performing model utilised FLAIR, T1, constitutional and disease non-imaging data, achieving an $R^2$ of 0.126, r of 0.355, and MAE of 0.584mmol/L (Figure 3).

A linear mixed-effect model predicting $R^2$ from the inclusion of T1/FLAIR, DWI, rsfMRI connectivity, disease, and psychology data for all 32 model input combinations for the 4 serology targets (128 models) found the inclusion of both disease (coefficient 0.014, 95% CI 0.008 to 0.021, p<0.001) and T1/FLAIR (coefficient 0.008, 95% CI 0.001 to 0.014, p=0.023) significantly advantageous to model performance (Figure 4). Conversely, rsfMRI connectivity was significantly associated with detrimental model performance (coefficient -0.025, 95% CI -0.032 to -0.019, p<0.001). There was no significant relationship between DWI and model performance (p=0.891).

The trade-off between computational time and performance

The total time required to train and optimize all models (both low and high resolution) was 188.589 P100 (16Gb) GPU days (4526.135 GPU hours). The total time required to train all 700 individual 64 x 64 x 64 resolution models was 146.281 GPU days (3510.734 GPU hours). The total time required to train the 25 individual 128 x 128 x 128 resolution models was 42.308 GPU days (1015.401 GPU hours).

For training 3D volumetric 64 x 64 x 64 imaging models, mean training time was 300.920 minutes (95% CI 279.316 to 322.524 minutes) (Figure 5). As expected, models that only included rsfMRI connectivity and/or non-imaging data took much less time to train (anywhere from just a few minutes for rsfMRI connectivity alone, to up to 123 minutes for constitutional, disease and psychology non-imaging data models. Models incorporating 3D volumetric imaging required far longer training times, up to a mean of 1412.750 minutes for predicting constitutional targets with rsfMRI connectivity, T1, FLAIR, and DWI. We cite training times for two reasons. First, they indicate the computational requirements for training uni- and/or multimodal deep models, including with multi-channel 3D imaging. Second, they show that the model performance reported here is unlikely to be trivially constrained by available compute, and more likely reflects the nature of the data and architectural limitations.

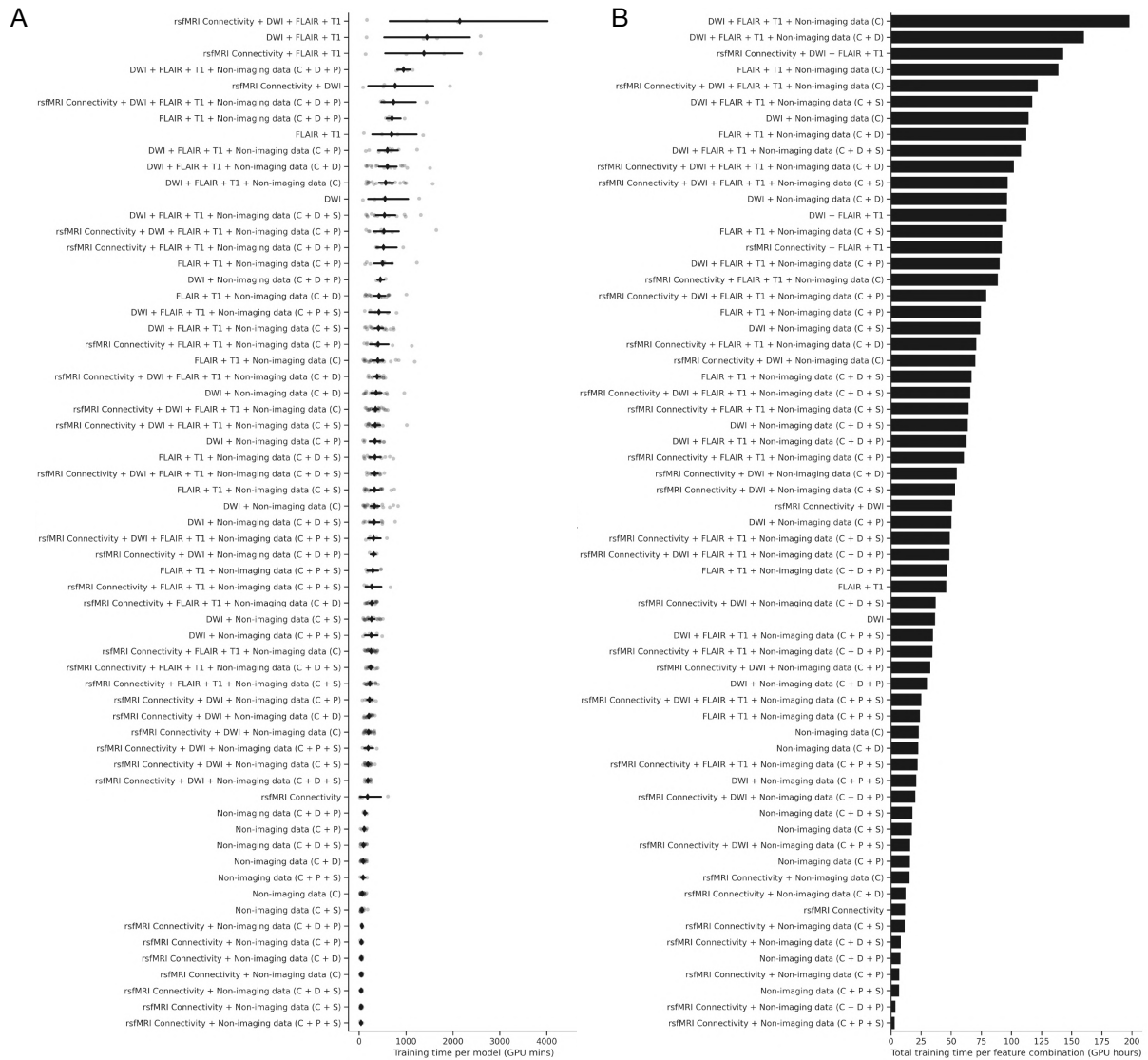

**Figure 5: Time costs in training large medical imaging models.** A) Strip plot illustrates training time taken per model in GPU minutes (x-axis) for all possible feature input combinations (y-axis). Grey points indicate individual models, with mean shown as a black diamond, and 95% confidence interval shown as a black line. B) Bar plot of total training time in GPU hours for all feature input combinations. Only 64 x 64 x 64 resolution models are shown here for visual simplicity.

Graph relationships of model performance

Finally, we created an undirected graph to visualise the similarities and differences of the model targets in terms of their predictability from different inputs (Figure 6). This showed that, whilst pairwise interrelation of participant features generally linked constitutional serological features, whereas psychological were highly interrelated (Figure 6A-B), pairwise interrelation linked by predictive fidelity revealed a segregation of constitutional features from those serological, with interrelation between features of disease and psychology, This graph is also available as a fully interactive, customizable, and downloadable HTML object (Supplementary Material). We also provide tabular data of performance metrics for all 700 models in the supplementary material, with all 700 individual model weights made open source.

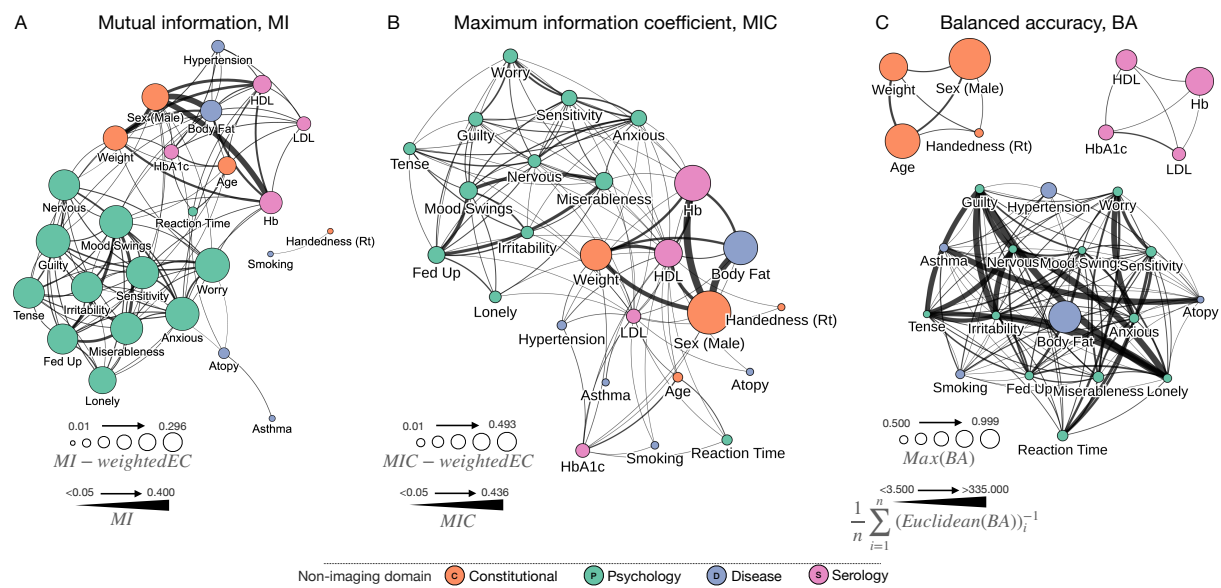

Figure 6: Static visual network analysis plots of feature relationships and model performances. A) Graph of target features, with nodes sized by mutual information (MI)-weighted eigenvector centrality (EC), and edges sized according to pairwise MI. Eigenvector centrality is a measure of influence of a node across a network. B) Graph of target features, with nodes sized by the maximum information coefficient (MIC)-weighted eigenvector centrality (EC), and edges sized according to the MIC. C) Graph of target features, with nodes sized by the maximum balanced accuracy across all models (BA), with edges sized according to the mean inverse Euclidean distance of all input combinations between each pair of targets. For all panels we depict the top 60% of edges for visualisation purposes. *Note that all graphs are made available as fully interactive and customizable HTML objects within the supplementary material.*

# Discussion

In the most comprehensive published analysis of its kind, we have quantified the individual-level predictability of 25 different constitutional, psychological, chronic disease, and serological characteristics, drawing on four different neuroimaging modalities spanning both structural and functional domains, and involving all possible combinations of features. The comparative performance of 700 models, trained over 189 GPU days (4526.135 GPU hours), with large-scale data from 23 810 individuals, casts light on the limits to prediction under a practically ideal modelling regime: large-scale data, state-of-the-art model architectures, and high-performance compute.

## Comparative legibility of biological features

Our study seeks to identify the limit on achievable individual-level predictive fidelity within the prevailing data regime. To do this, we are obliged to use state-of-the-art methods, for any shortfall in performance could otherwise be attributed to a remediable deficiency in the model architecture. Our models of constitutional features set new state-of-the-art benchmarks for age and sex prediction from neuroimaging data[81,82]. Though models combining T1, FLAIR and DWI performed best, achieving balanced accuracy of 99.7% for sex and MAE of 2.048 years for age, even rsfMRI connectivity alone achieved a balanced accuracy of 91.5% for sex, a new state-of-the-art for non-structural imaging[84]. Equally, a MAE of 7.04Kg indicates remarkably high fidelity for weight, a hitherto under-explored task here rendered maximally challenging by the exclusion of all non-brain tissue prior to modelling. These results show that our model architectures and overall analytic pipeline faithfully reflect the highest standards in the field, supporting a rigorous test of the current limits on the individual predictability of other features. That the remaining constitutional feature—handedness, best predicted using rsfMRI connectivity data alone—achieved a balanced accuracy of only 57.7% suggests the difficulty here does not arise from inadequate implementation of current technology. It is striking, even accounting for class imbalance, that handedness is individually so poorly predictable given the magnitude of population-level differences in the organisation of the brain[85-87].

Psychological targets showed equally limited predictability, maximal from disease data. The addition of any neuroimaging, whether structural or functional, generally offered no material benefit. Of the 12 psychological targets, only the propensity to worry and reaction time showed some effect from neuroimaging, but only in the context of low overall performance. This is again strikingly at odds with population-level observations, where marked group-level differences are often reported, but individual level predictability is relatively low.

Disease targets, focused on common conditions without gross, diagnostic imaging changes, were also poorly predictable at the individual level, with serology the strongest predictor. Only percentage body fat benefited from imaging (DWI, FLAIR and T1), achieving MAE of 2.65%, but not far from age, sex, and weight alone (3.01%).

Serology was best predicted by disease, followed by T1/FLAIR structural imaging. Haemoglobin offered the best—and most multimodal—performance, drawing on DWI, FLAIR, T1, constitutional, disease, and psychological data (MAE 0.629 g/dl), but its prediction is likely to lean on covarying sex and age. Although we did not explicitly test the question, the difference from constitutional data alone (MAE 0.676g/dl) is likely explained by global differences in the MRI signal.

These performance figures were essentially invariant to modelled image resolution. Across all 25 model targets, smoking was the only one to demonstrate an increase in model fidelity when training at a higher imaging resolution (128 x 128 x 128). The gain, however, was marginal: 58.1% rising to 59.2%. Under current data and architectural regimes, meaningful improvements in performance are unlikely to be achieved merely by increasing image resolution.

The limits of individual prediction

Our analysis shows that whereas constitutional characteristics—age, sex, and weight—are highly predictable from neuroimaging, psychology, chronic illness, and serological characteristics are not. Crucially, *comparative* differences in predictability are extraordinarily high, suggesting that with currently practicable models, the limits are primarily set by the fundamental informativity of imaging signals. Substantially higher individual resolving power will require either a radical 'regime change'—in terms of volumes of data, model expressivity, and compute—or new investigational methods.

Three implications are foremost. First, imaging-based clinical decision-support systems with cognitive or behavioural targets—*operating in the absence of overt changes on imaging*—will likely continue to be plagued by underperformance, especially when deployed in real-world scenarios. If fundamental psychological characteristics are illegible under a modelling regime far more conducive to success than clinical realities ordinarily permit, the prospects of such endeavours seem dim. Performance in specific clinical populations inadequately sampled by population-based studies such as UK Biobank[1], especially those with overtly abnormal imaging, may well be higher, but the bar is clearly set high.

Second, population-level mechanistic models of cognition and behaviour that seek to ground theories of brain function in terms of normal structural and resting-state functional features will likely leave most individual variability unexplained[14]. Though many in the field consider explanatory and predictive power to be decoupled, the claim to fidelity of a theory contradicted by most instances it describes is bound to strike a disinterested observer as insecure. Were residual variability truly random when inspected at finer scales of observation, with more powerful methods such as intracranial recording, then a case could be made that the observed stochasticity is more aleatoric than epistemic. But such studies near-universally reveal complex structure a suitably expressive model could conceivably capture.

Third, the manifest difficulty of prediction mandates obtaining not just the largest but also the widest possible data support, operating multi-modally, with resilience to the missingness and noise corruption ubiquitous in the real world. This implies a decisive shift not only away from simple models but also from unimodal discriminative models of any complexity. Only multimodal generative models drawing on all available data, complete and incomplete, could adequately corroborate the belief any observed underperformance is irremediable without new or higher quality known biological signals[50].

Limitations

Our inferences are supported by the largest and most comprehensive evaluation of its kind. Nonetheless, the data and modelling context give rise to an array of limitations.

First, we chose to model 25 non-imaging characteristics from a far wider range of data available in UK Biobank[1,21,22,88]. A modest number is inevitable where, as here, the objective is to probe the effect of multiple *combinations* of characteristics, with large-scale data and comprehensive model optimization. Indeed, the task of training 700 independent deep learning models of this kind is already onerous enough not to have been previously attempted in this domain. The choice of as wide a range of characteristics as was feasible for the compute at our disposal is deliberate, for it allows us to evaluate the predictive contribution of signals distributed across *multiple* characteristics. Not doing this could have raised the possibility that a given unimodal signal may be present, but camouflaged by multi-modal contextual modulation. Note that each individual characteristic is comprehensively modelled in any event, and the multimodal perspective is an addition, not a substitution.

Second, it is conceivable that architectural—as opposed to hyper-parameter—tuning may have obtained better performance for any one individual target. But our objective here was to standardize the model architecture across all possible targets and feature combinations, creating a general-purpose prediction pipeline that enables a fair comparison of the distinct contribution of each input. Moreover, reasoning that non-constitutional targets may require large numbers of parameters, we employed more flexible architectures than current age and sex classifiers[80,81]. Note the flexibility was not such as to induce overfitting on constitutional targets, so potential overfitting on other targets is not explained by excessive overparameterisation.

Third, it is possible that aspects of the processing upstream of the predictive modelling may have an impact on performance. Such variabilities are common across neuroimaging research (and indeed form a focus of other research groups[89-91]). The same is true for our use of widely available resting-state (as opposed to task-related) fMRI data, or use of other complimentary imaging modalities such as EEG[92], where further improvement in model performance is foreseeable. But here we adopt common pre-processing practices[21] and parcellations[31,38], and it seems unlikely that the striking differences in performance observed here, especially with state-of-the-art constitutional performance, are thereby explained. Task-based fMRI may reveal greater predictive fidelity in some tasks, especially when compared to resting state. We however deliberately did not include such data for our priority was to maximise generalizability, for the nature of a task-based functional scan could vary greatly across any research study or imaging site.

Fourth, UK Biobank's cohort, though peerlessly large for phenotyping of this richness, is explicitly limited to an older age group and implicitly limited by the time and dispositional demands of participation. The observed variations of the features of interest are, however, both generous and comparable to those likely to obtain in real-world contexts. Equally, the quality and instrumental homogeneity of the source data may theoretically be exceeded elsewhere, even if there is no superior study currently in progress. The critical question we address in this article is the limit on individual-level fidelity imposed by current data regimes, a task that necessitates the largest and richest available unselected dataset, of which UK Biobank is internationally the leading example. While generalisability is definitionally impossible to assure universally, this is as close as anyone could plausibly get at present.

Fifth, the predictive targets are deliberately chosen to exclude those diagnosable from imaging (e.g., acute stroke), for the task then becomes one of recognition rather than prediction[93]. Although chronic diseases of high prevalence are included, our focus is on characteristics common enough to span the normal/abnormal divide, at least in statistical terms. This focus reflects the scale of potential population-level benefit in illuminating individual-level patterns of the underlying substrates and processes as reflected in the imaged brain.

Finally, psychological characteristics can only be imperfectly captured by test instruments whose reliability is bound to vary, both across characteristics and datasets. Better tests may, of course, provide targets with higher achievable fidelity. But the variations in observed reliability[94] are not so large as to trivially explain the striking differences in comparative predictability here, and the chosen tests are known to correlate reasonably well with other measures[95,96].

Conclusion

In the largest study of its kind, involving 700 models trained on a comprehensive set of combinations of 25 target biological features, across multiple domains and 23 810 unique participants, we have quantified the individual-level legibility of the human brain. Determining the *comparative* predictability of different targets from each other and from multimodal brain imaging, under the current practical maximum of data quality, algorithmic felicity, and computational resource, we set out to answer a key strategic question: is actionable individual-level predictive fidelity plausibly achievable under current data regimes, or is a radical change necessary? The striking difference in observed comparative predictability suggests the latter, interpretative limitations notwithstanding. If predictive systems are to achieve the individual-level fidelity clinical utility demands, and if mechanistic models are to capture enough variability in the population to be persuasively generalizable, regime change is now unavoidable.

# References

1   Littlejohns, T. J. *et al.* The UK Biobank imaging enhancement of 100,000 participants: rationale, data collection, management and future directions. *Nat Commun* **11**, 2624 (2020). https://doi.org:10.1038/s41467-020-15948-9

2   Bethlehem, R. A. I. *et al.* Brain charts for the human lifespan. *Nature* **604**, 525-533 (2022). https://doi.org:10.1038/s41586-022-04554-y

3   Elliott, L. T. *et al.* Genome-wide association studies of brain imaging phenotypes in UK Biobank. *Nature* **562**, 210-216 (2018). https://doi.org:10.1038/s41586-018-0571-7

4   Wang, C. *et al.* Phenotypic and genetic associations of quantitative magnetic susceptibility in UK Biobank brain imaging. *Nat Neurosci* **25**, 818-831 (2022). https://doi.org:10.1038/s41593-022-01074-w

5   Bazinet, V. *et al.* Assortative mixing in micro-architecturally annotated brain connectomes. *Nature Communications* **14**, 2850 (2023). https://doi.org:10.1038/s41467-023-38585-4

6   Hansen, J. Y. *et al.* Mapping neurotransmitter systems to the structural and functional organization of the human neocortex. *Nature Neuroscience* **25**, 1569-1581 (2022). https://doi.org:10.1038/s41593-022-01186-3

7   Suárez, L. E., Markello, R. D., Betzel, R. F. & Misic, B. Linking Structure and Function in Macroscale Brain Networks. *Trends in Cognitive Sciences* **24**, 302-315 (2020). https://doi.org:https://doi.org/10.1016/j.tics.2020.01.008

8   Honey, C. J. *et al.* Predicting human resting-state functional connectivity from structural connectivity. *Proceedings of the National Academy of Sciences* **106**, 2035-2040 (2009). https://doi.org:10.1073/pnas.0811168106

9   Fischl, B. *et al.* Cortical Folding Patterns and Predicting Cytoarchitecture. *Cerebral Cortex* **18**, 1973-1980 (2008). https://doi.org:10.1093/cercor/bhm225

10  Thomas Yeo, B. T. *et al.* The organization of the human cerebral cortex estimated by intrinsic functional connectivity. *Journal of Neurophysiology* **106**, 1125-1165 (2011). https://doi.org:10.1152/jn.00338.2011

11  Hansen, J. Y. *et al.* Mapping gene transcription and neurocognition across human neocortex. *Nature Human Behaviour* **5**, 1240-1250 (2021). https://doi.org:10.1038/s41562-021-01082-z

12  Marek, S. *et al.* Reproducible brain-wide association studies require thousands of individuals. *Nature* **603**, 654-660 (2022). https://doi.org:10.1038/s41586-022-04492-9

13   Finn, E. S. *et al.* Functional connectome fingerprinting: identifying individuals using patterns of brain connectivity. *Nat Neurosci* **18**, 1664-1671 (2015). https://doi.org:10.1038/nn.4135

14   Bzdok, D., Engemann, D. & Thirion, B. Inference and Prediction Diverge in Biomedicine. *Patterns* **1**, 100119 (2020). https://doi.org:https://doi.org/10.1016/j.patter.2020.100119

15   Wu, J., Li, J., Eickhoff, S. B., Scheinost, D. & Genon, S. The challenges and prospects of brain-based prediction of behaviour. *Nature Human Behaviour* **7**, 1255-1264 (2023). https://doi.org:10.1038/s41562-023-01670-1

16   LeCun, Y., Bengio, Y. & Hinton, G. Deep learning. *Nature* **521**, 436-444 (2015). https://doi.org:10.1038/nature14539

17   Goodfellow, I., Bengio, Y. & Courville, A. *Deep Learning*. (MIT Press, 2017).

18   Richards, B. A. *et al.* A deep learning framework for neuroscience. *Nature Neuroscience* **22**, 1761-1770 (2019). https://doi.org:10.1038/s41593-019-0520-2

19   Szucs, D. & Ioannidis, J. P. Sample size evolution in neuroimaging research: An evaluation of highly-cited studies (1990-2012) and of latest practices (2017-2018) in high-impact journals. *Neuroimage* **221**, 117164 (2020). https://doi.org:10.1016/j.neuroimage.2020.117164

20   Schulz, M.-A. *et al.* Different scaling of linear models and deep learning in UKBiobank brain images versus machine-learning datasets. *Nature Communications* **11**, 4238 (2020). https://doi.org:10.1038/s41467-020-18037-z

21   Alfaro-Almagro, F. *et al.* Image processing and Quality Control for the first 10,000 brain imaging datasets from UK Biobank. *Neuroimage* **166**, 400-424 (2018). https://doi.org:10.1016/j.neuroimage.2017.10.034

22   Sudlow, C. *et al.* UK Biobank: An Open Access Resource for Identifying the Causes of a Wide Range of Complex Diseases of Middle and Old Age. *PLOS Medicine* **12**, e1001779 (2015). https://doi.org:10.1371/journal.pmed.1001779

23   Pedregosa, F., Varoquaux, G. & Gramfort, A. Scikit-learn: Machine Learning in Python. *Journal of Machine Learning Research* **12**, 2825-2830 (2011).

24   Reshef, D. N. *et al.* Detecting novel associations in large data sets. *Science* **334**, 1518-1524 (2011). https://doi.org:10.1126/science.1205438

25   Smith, S. M. *et al.* Advances in functional and structural MR image analysis and implementation as FSL. *Neuroimage* **23 Suppl 1**, S208-219 (2004). https://doi.org:10.1016/j.neuroimage.2004.07.051

26   Smith, S. M. Fast robust automated brain extraction. *Hum Brain Mapp* **17**, 143-155 (2002). https://doi.org:10.1002/hbm.10062

27	Jenkinson, M., Bannister, P., Brady, M. & Smith, S. Improved optimization for the robust and accurate linear registration and motion correction of brain images. *Neuroimage* **17**, 825-841 (2002).

28	Grabner, G. *et al.* in *Medical Image Computing and Computer-Assisted Intervention – MICCAI 2006.* (eds Rasmus Larsen, Mads Nielsen, & Jon Sporring) 58-66 (Springer Berlin Heidelberg).

29	Smith, S. M. *et al.* Tract-based spatial statistics: voxelwise analysis of multi-subject diffusion data. *Neuroimage* **31**, 1487-1505 (2006). https://doi.org:10.1016/j.neuroimage.2006.02.024

30	Abraham, A. *et al.* Machine learning for neuroimaging with scikit-learn. *Front Neuroinform* **8**, 14 (2014). https://doi.org:10.3389/fninf.2014.00014

31	Glasser, M. F. *et al.* A multi-modal parcellation of human cerebral cortex. *Nature* **536**, 171-178 (2016). https://doi.org:10.1038/nature18933

32	Bullmore, E. & Sporns, O. Complex brain networks: graph theoretical analysis of structural and functional systems. *Nat Rev Neurosci* **10**, 186-198 (2009). https://doi.org:10.1038/nrn2575

33	Zalesky, A., Fornito, A. & Bullmore, E. T. Network-based statistic: identifying differences in brain networks. *Neuroimage* **53**, 1197-1207 (2010). https://doi.org:10.1016/j.neuroimage.2010.06.041

34	Ruffle, J. K. *et al.* The autonomic brain: Multi-dimensional generative hierarchical modelling of the autonomic connectome. *Cortex* **143**, 164-179 (2021). https://doi.org:https://doi.org/10.1016/j.cortex.2021.06.012

35	Logothetis, N. K., Pauls, J., Augath, M., Trinath, T. & Oeltermann, A. Neurophysiological investigation of the basis of the fMRI signal. *Nature* **412**, 150-157 (2001). https://doi.org:10.1038/35084005

36	Raamana, P. R., Weiner, M. W., Wang, L. & Beg, M. F. Thickness network features for prognostic applications in dementia. *Neurobiology of Aging* **36**, S91-S102 (2015). https://doi.org:https://doi.org/10.1016/j.neurobiolaging.2014.05.040

37	Thiebaut de Schotten, M., Foulon, C. & Nachev, P. Brain disconnections link structural connectivity with function and behaviour. *Nat Commun* **11**, 5094 (2020). https://doi.org:10.1038/s41467-020-18920-9

38	Glasser, M. F. *et al.* The Human Connectome Project's neuroimaging approach. *Nat Neurosci* **19**, 1175-1187 (2016). https://doi.org:10.1038/nn.4361

39	Benjamini, Y. & Yekutieli, D. False Discovery Rate–Adjusted Multiple Confidence Intervals for Selected Parameters. *Journal of the American Statistical Association* **100**, 71-81 (2005). https://doi.org:10.1198/016214504000001907

40  Peixoto, T. P. Inferring the mesoscale structure of layered, edge-valued, and time-varying networks. *Physical Review E* **92**, 042807 (2015). https://doi.org:10.1103/PhysRevE.92.042807

41  Cipolotti, L. *et al.* Graph lesion-deficit mapping of fluid intelligence. *Brain* **146**, 167-181 (2022). https://doi.org:https://doi.org/10.1101/2022.07.28.501722

42  Peixoto, T. P. Efficient Monte Carlo and greedy heuristic for the inference of stochastic block models. *Physical Review E* **89**, 012804 (2014). https://doi.org:10.1103/PhysRevE.89.012804

43  Peixoto, T. P. Entropy of stochastic blockmodel ensembles. *Physical Review E* **85**, 056122 (2012). https://doi.org:10.1103/PhysRevE.85.056122

44  Ruffle, J. K. *et al.* Brain tumour genetic network signatures of survival. *Brain* (2023).

45  Bakas, S. *et al.* Identifying the Best Machine Learning Algorithms for Brain Tumor Segmentation, Progression Assessment, and Overall Survival Prediction in the BRATS Challenge. *ArXiv* **abs/1811.02629** (2018).

46  Isensee, F., Jaeger, P. F., Kohl, S. A. A., Petersen, J. & Maier-Hein, K. H. nnU-Net: a self-configuring method for deep learning-based biomedical image segmentation. *Nat Methods* **18**, 203-211 (2021). https://doi.org:10.1038/s41592-020-01008-z

47  Jonsson, B. A. *et al.* Brain age prediction using deep learning uncovers associated sequence variants. *Nat Commun* **10**, 5409 (2019). https://doi.org:10.1038/s41467-019-13163-9

48  Ruffle, J., K., Mohinta, S., Gray, R., Hyare, H. & Nachev, P. Brain tumour segmentation with incomplete imaging data. *Brain Commun* (2023). https://doi.org:10.1093/braincomms/fcad118

49  Consortium, T. M. Project MONAI. *Zenodo* (2020). https://doi.org:https://doi.org/10.5281/zenodo.4323059

50  Pinaya, W. H. L. *et al.* Generative AI for Medical Imaging: extending the MONAI Framework. *arXiv e-prints*, arXiv:2307.15208 (2023). https://doi.org:10.48550/arXiv.2307.15208

51  Paszke, A. *et al.* PyTorch: An Imperative Style, High-Performance Deep Learning Library. *NeurIPS* (2019).

52  Ioffe, S. & Szegedy, C. in *Proceedings of the 32nd International Conference on Machine Learning* Vol. 37  (eds Bach Francis & Blei David) 448--456 (PMLR, Proceedings of Machine Learning Research, 2015).

53  Hendrycks, D. & Gimpel, K. Gaussian Error Linear Units (GELUs). *arXiv* (2016).

54  Srivastava, N., Hinton, G., Krizhevsky, A., Sutskever, I. & Salakhutdinov, R. Dropout: a simple way to prevent neural networks from overfitting. *J. Mach. Learn. Res.* **15**, 1929–1958 (2014).

55  Krizhevsky, A., Sutskever, I. & Hinton, G. ImageNet Classification with Deep Convolutional Neural Networks. *Advances in Neural Information Processing Systems* **25** (2012).

56  He, K., Zhang, X., Ren, S. & Sun, J. Deep Residual Learning for Image Recognition. *2016 IEEE Conference on Computer Vision and Pattern Recognition (CVPR)*, 770-778 (2016).

57  Yamaguchi, K., Sakamoto, K., Akabane, T. & Fujimoto, Y. A Neural Network for Speaker-Independent Isolated Word Recognition. *ICSLP 90*, 1077-1080 (1990).

58  Heinz, S. *A performance benchmark of Google AutoML Vision using Fashion-MNIST*, <https://towardsdatascience.com/a-performance-benchmark-of-google-automl-vision-using-fashion-mnist-a9bf8fc1c74f> (2018).

59  Benchmarks.AI. *MNIST*, <https://benchmarks.ai/mnist> (2021).

60  Ahn, H. & Yim, C. Convolutional Neural Networks Using Skip Connections with Layer Groups for Super-Resolution Image Reconstruction Based on Deep Learning. *Applied Sciences* **10**, 1959 (2020).

61  Kingma, D. P. & Ba, J. Adam: A Method for Stochastic Optimization. *arXiv* **1412.6980** (2017).

62  Cortes, C., Mohri, M. & Rostamizadeh, A. L2 Regularization for Learning Kernels. arXiv:1205.2653 (2012). <https://ui.adsabs.harvard.edu/abs/2012arXiv1205.2653C>.

63  Trimarchi, D. *Confusion Matrix*, <https://github.com/DTrimarchi10/confusion_matrix> (2019).

64  Varoquaux, G. & Colliot, O. in *Machine Learning for Brain Disorders*  (ed Olivier Colliot) 601-630 (Springer US, 2023).

65  Poldrack, R. A., Huckins, G. & Varoquaux, G. Establishment of Best Practices for Evidence for Prediction: A Review. *JAMA Psychiatry* **77**, 534-540 (2020). https://doi.org:10.1001/jamapsychiatry.2019.3671

66  Pinheiro, J. *et al.* Package 'nlme'. *cran*, 1-328 (2022).

67  Wickham, H. *et al.* Welcome to the Tidyverse. *Journal of Open Source Software* **4** (2019).

68  Peixoto, T. P. The graph-tool python library. *figshare* (2014). https://doi.org:10.6084/m9.figshare.1164194

69  Haas, R. *gravis*, <https://github.com/robert-haas/gravis> (2021).

70  Tange, O. GNU Parallel - The Command-Line Power Tool. *The USENIX Magazine*, 42-47 (2011).

71  Hunter, J. D. Matplotlib: A 2D Graphics Environment. *Computing in Science & Engineering* **9**, 90-95 (2007). https://doi.org:10.1109/MCSE.2007.55

72  Albanese, D., Riccadonna, S., Donati, C. & Franceschi, P. A practical tool for maximal information coefficient analysis. *GigaScience* **7** (2018). https://doi.org:10.1093/gigascience/giy032

73  Brett, Matthew, Markiewicz & Hanke, C. nipy/nibabel: 3.2.1 (Version 3.2.1). *Zenodo* (2020). https://doi.org:http://doi.org/10.5281/zenodo.4295521

74  Harris, C. R. *et al.* Array programming with NumPy. *Nature* **585**, 357-362 (2020). https://doi.org:10.1038/s41586-020-2649-2

75  Reback, J., McKinney, W. & jbrockmendel. pandas-dev/pandas: Pandas 1.0.3 (Version v1.0.3). *Zenodo* (2020). https://doi.org:http://doi.org/10.5281/zenodo.3715232

76  Virtanen, P. *et al.* SciPy 1.0: fundamental algorithms for scientific computing in Python. *Nature Methods* **17**, 261-272 (2020). https://doi.org:10.1038/s41592-019-0686-2

77  Waskom, M. & Seaborn_Development_Team. seaborn. *Zenodo* (2020). https://doi.org:10.5281/zenodo.592845

78  Seabold, S. & Perktold, J. Statsmodels: Econometric and Statistical Modeling with Python. *Proc of the 9th Python in science conference* (2010).

79  Developers, N. *CUDA Toolkit 11.0*, <https://developer.nvidia.com/cuda-11.0-download-archive> (2021).

80  Cole, J. H. Multimodality neuroimaging brain-age in UK biobank: relationship to biomedical, lifestyle, and cognitive factors. *Neurobiol Aging* **92**, 34-42 (2020). https://doi.org:10.1016/j.neurobiolaging.2020.03.014

81  Peng, H., Gong, W., Beckmann, C. F., Vedaldi, A. & Smith, S. M. Accurate brain age prediction with lightweight deep neural networks. *Medical Image Analysis* **68**, 101871 (2021). https://doi.org:https://doi.org/10.1016/j.media.2020.101871

82  Gong, W., Beckmann, C. F., Vedaldi, A., Smith, S. M. & Peng, H. Optimising a Simple Fully Convolutional Network for Accurate Brain Age Prediction in the PAC 2019 Challenge. *Front Psychiatry* **12**, 627996 (2021). https://doi.org:10.3389/fpsyt.2021.627996

83  Peixoto, T. P. Nonparametric weighted stochastic block models. *Physical Review E* **97**, 012306 (2018). https://doi.org:10.1103/PhysRevE.97.012306

84  Leming, M. & Suckling, J. Deep learning for sex classification in resting-state and task functional brain networks from the UK Biobank. *Neuroimage* **241**, 118409 (2021). https://doi.org:10.1016/j.neuroimage.2021.118409

85  Sha, Z. *et al.* Handedness and its genetic influences are associated with structural asymmetries of the cerebral cortex in 31,864 individuals. *Proc Natl Acad Sci U S A* **118** (2021). https://doi.org:10.1073/pnas.2113095118

86  Chormai, P. *et al.* Machine learning of large-scale multimodal brain imaging data reveals neural correlates of hand preference. *Neuroimage* **262**, 119534 (2022). https://doi.org:10.1016/j.neuroimage.2022.119534

87  Good, C. D. *et al.* Cerebral asymmetry and the effects of sex and handedness on brain structure: a voxel-based morphometric analysis of 465 normal adult human brains. *Neuroimage* **14**, 685-700 (2001). https://doi.org:10.1006/nimg.2001.0857

88  Bycroft, C. *et al.* The UK Biobank resource with deep phenotyping and genomic data. *Nature* **562**, 203-209 (2018). https://doi.org:10.1038/s41586-018-0579-z

89  Fusar-Poli, P. *et al.* Effect of image analysis software on neurofunctional activation during processing of emotional human faces. *J Clin Neurosci* **17**, 311-314 (2010). https://doi.org:10.1016/j.jocn.2009.06.027

90  Haddad, E. *et al.* Multisite test-retest reliability and compatibility of brain metrics derived from FreeSurfer versions 7.1, 6.0, and 5.3. *Hum Brain Mapp* **44**, 1515-1532 (2023). https://doi.org:10.1002/hbm.26147

91  Zhou, X. *et al.* Choice of Voxel-based Morphometry processing pipeline drives variability in the location of neuroanatomical brain markers. *Communications Biology* **5**, 913 (2022). https://doi.org:10.1038/s42003-022-03880-1

92  Chowdhury, M. S. N. *et al.* Deep Neural Network for Visual Stimulus-Based Reaction Time Estimation Using the Periodogram of Single-Trial EEG. *Sensors (Basel)* **20** (2020). https://doi.org:10.3390/s20216090

93  Farazi, H. & Nogga, J. Semantic Prediction: Which One Should Come First, Recognition or Prediction?  , arXiv:2110.02829 (2021). <https://ui.adsabs.harvard.edu/abs/2021arXiv211002829F>.

94  Fawns-Ritchie, C. & Deary, I. J. Reliability and validity of the UK Biobank cognitive tests. *PLOS ONE* **15**, e0231627 (2020). https://doi.org:10.1371/journal.pone.0231627

95  Wu, J. *et al.* Cross-cohort replicability and generalizability of connectivity-based psychometric prediction patterns. *NeuroImage* **262**, 119569 (2022). https://doi.org:https://doi.org/10.1016/j.neuroimage.2022.119569

96  He, T. *et al.* Deep neural networks and kernel regression achieve comparable accuracies for functional connectivity prediction of behavior and demographics. *NeuroImage* **206**, 116276 (2020). https://doi.org:https://doi.org/10.1016/j.neuroimage.2019.116276

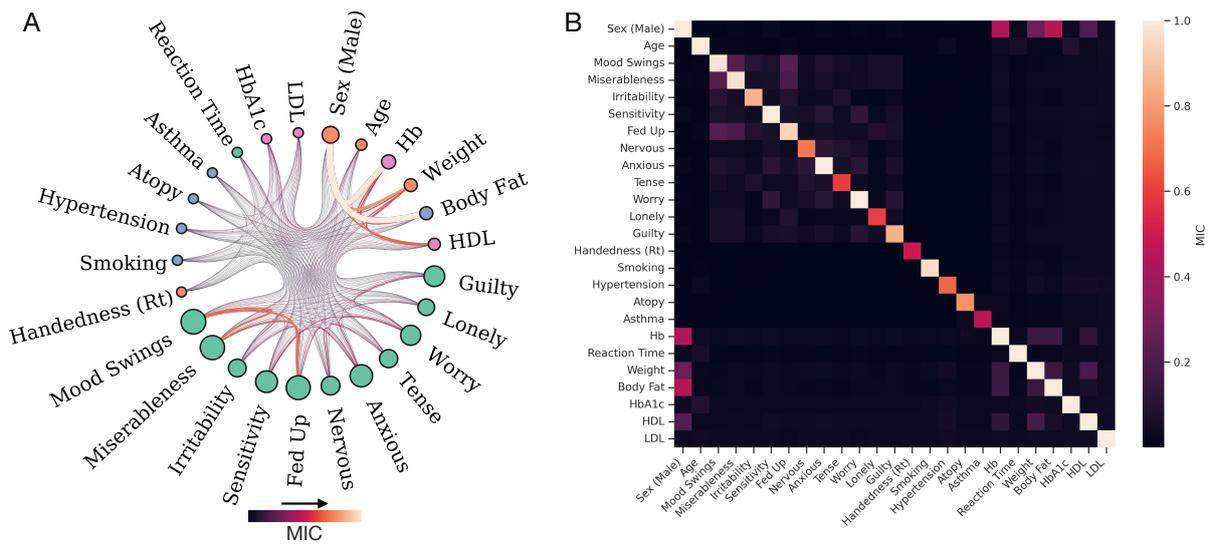

**Supplementary Figure 1: Target interrelation.** A) Stochastic block model illustrating feature relation by MIC. B) Heatmap of MIC relational values. The process identifies that participant sex holds some degree of informational relationship between their haemoglobin (Hb) concentration, weight, percentage body fat, and high-density lipoprotein (HDL) concentration. Similarly, psychological metrics generally seem to have weak (but nonetheless present) informational relationships to each other, and the same is true between serological variables. Notably the MIC matrix forms the justification for excluding metadata from the same domain as the target for any training run.

| Target (Domain) | Biobank Data ID |
|---|---|
| Sex (C) | https://biobank.ctsu.ox.ac.uk/crystal/coding.cgi?id=9 |
| Age (C) | https://biobank.ctsu.ox.ac.uk/crystal/field.cgi?id=21003 |
| Weight (C) | https://biobank.ctsu.ox.ac.uk/crystal/field.cgi?id=21002 |
| Handedness (C) | http://biobank.ctsu.ox.ac.uk/crystal/field.cgi?id=1707 |
| Mood Swings (P) | https://biobank.ctsu.ox.ac.uk/crystal/field.cgi?id=1920 |
| Miserableness (P) | https://biobank.ctsu.ox.ac.uk/crystal/field.cgi?id=1930 |
| Irritability (P) | https://biobank.ctsu.ox.ac.uk/crystal/field.cgi?id=1940 |
| Sensitivity (P) | https://biobank.ctsu.ox.ac.uk/crystal/field.cgi?id=1950 |
| Fed Up (P) | https://biobank.ctsu.ox.ac.uk/crystal/field.cgi?id=1960 |
| Nervous (P) | https://biobank.ctsu.ox.ac.uk/crystal/field.cgi?id=1970 |
| Anxious (P) | https://biobank.ctsu.ox.ac.uk/crystal/field.cgi?id=1980 |
| Tense (P) | https://biobank.ctsu.ox.ac.uk/crystal/field.cgi?id=1990 |
| Worry (P) | https://biobank.ctsu.ox.ac.uk/crystal/field.cgi?id=2000 |
| Lonely (P) | https://biobank.ctsu.ox.ac.uk/crystal/field.cgi?id=2020 |
| Guilty (P) | https://biobank.ctsu.ox.ac.uk/crystal/field.cgi?id=2030 |
| Reaction Time (ms) (P) | https://biobank.ctsu.ox.ac.uk/crystal/field.cgi?id=20023 |
| Smoking (D) | https://biobank.ctsu.ox.ac.uk/crystal/field.cgi?id=20116 |
| Hypertension (D) | https://biobank.ctsu.ox.ac.uk/crystal/coding.cgi?id=6 |
| Atopy (D) | https://biobank.ctsu.ox.ac.uk/crystal/field.cgi?id=3761 |
| Asthma (D) | https://biobank.ctsu.ox.ac.uk/crystal/field.cgi?id=3786 |
| Body Fat (%) (D) | https://biobank.ctsu.ox.ac.uk/crystal/field.cgi?id=23099 |
| Hb (g/dl) (S) | https://biobank.ctsu.ox.ac.uk/crystal/field.cgi?id=30020 |
| HbA1c (mmol/mol) (S) | http://biobank.ctsu.ox.ac.uk/crystal/field.cgi?id=30750 |
| HDL (mmol/L) (S) | http://biobank.ctsu.ox.ac.uk/crystal/field.cgi?id=30760 |
| LDL (mmol/L) (S) | https://biobank.ctsu.ox.ac.uk/crystal/field.cgi?id=30780 |

**Supplementary Table 1: UK Biobank Data-Fields.**

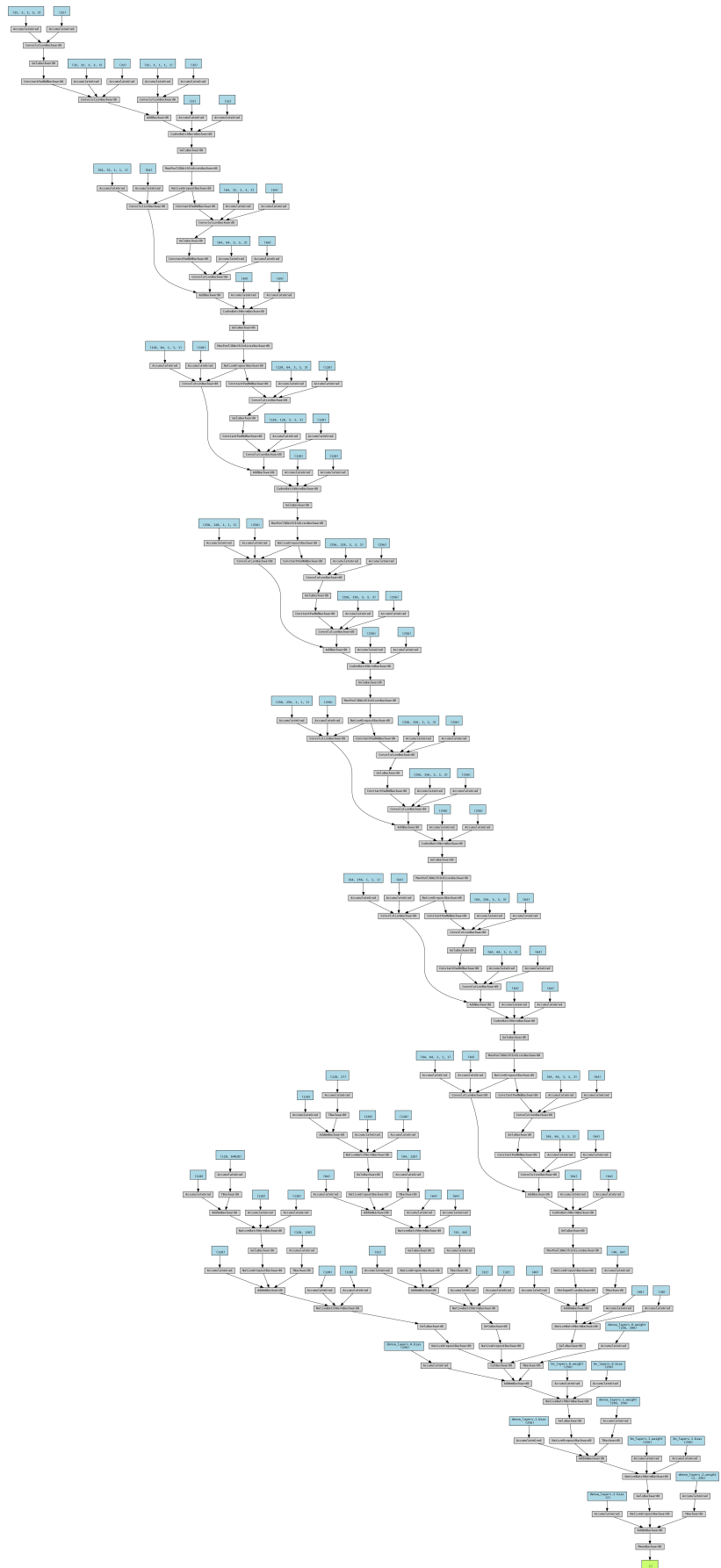

Supplementary Figure 2: Model architectures. Model architectures for the 3D convolutional neural network (CNN), metadata multilayer perceptron (MLP), rsfMRI connectivity MLP, and final prediction MLP.